\documentclass{aa} 
\usepackage{graphicx}
\usepackage{longtable}
\begin{document}
   \title{The core fundamental plane of B2 radio galaxies}


   \author{D. Bettoni
          \inst{1}
          \and
          R. Falomo\inst{1} \and P. Parma\inst{2} \and 
           H. de Ruiter \inst{3,2} \and R. Fanti\inst{2}
          }

   \offprints{D. Bettoni}

   \institute{INAF - Osservatorio Astronomico di Padova,
              Vicolo dell'Osservatorio, 5 -35122 Padova-ITALY\\
              \email{daniela.bettoni@oapd.inaf.it, renato.falomo@oapd.inaf.it}
         \and INAF - IRA Bologna, Italy \\
	     \email{paola.parma@ira.inaf.it,roberto.fanti@ira.inaf.it}
	 \and 
	 INAF-Osservatorio Astronomico di  Bologna, 
	 via Ranzani, 1 - 40127 Bologna, Italy  \\ 
          \email{hans.deruiter@ira.inaf.it}   }

   \date{Received...; accepted...}


  \abstract
   {The photometric, structural and kinematical  properties of the centers of elliptical 
   galaxies, harbour important information of the formation history of the galaxies. 
   In the case of non active elliptical galaxies these properties are linked in 
a way that surface brightness, break radius and velocity dispersion of the core 
lie on a fundamental plane similar to that found for their global properties.}
   {We construct the Core Fundamental Plane (CFP) for a sizeable sample of low redshift radio 
galaxies and compare it with that of non radio ellipticals.}
   {To pursue this aim we combine  data obtained from high resolution HST images with medium resolution optical 
   spectroscopy to derive the photometric and kinematic properties of $\sim$40 low redshift radio galaxies. }
   {We find that the CFPs of radio galaxies is indistinguishable from that 
   defined by non radio elliptical galaxies of similar luminosity. 
   The characteristics of the CFP of radio galaxies are also consistent (same slope) with those of the
Fundamental Plane (FP) derived from the global  properties of radio  (and non radio) 
elliptical galaxies. The similarity of CFP and FP for radio and non radio ellipticals suggests that the active phase of these galaxies has minimal effects for the structure of the galaxies.}
   {}

   \maketitle
%

\section{Introduction}

The properties of the centers of massive elliptical galaxies are of strategic importance for the
understanding of the complex processes of galaxy formation. The centers represent the bottom of the
potential well of the galaxy, host massive black holes and provide a record of the past history of
the galaxies. Until a decade ago the properties of the center of galaxies were very little studied
because of insufficient spatial resolution of available instrumentation. 
Only with the use of Hubble Space Telescope, and using future
large ground based telescope such a study become feasible.


In the pioneering study on the nuclear properties of nearby early type galaxies (carried out with
HST images) \cite{faber} were able to probe the inner regions of a number of nearby ellipticals. They point out
that the inner luminosity profiles can be parameterized by a core (or break) radius $R_b$ and a
characteristic surface brightness within the  break radius $<\mu_b>$. 
They also showed that these parameters ($R_b$,
$<\mu_b>$) can be combined to form 
a Core Fundamental Plane (CFP) which is 
analogous to the one found for the global properties of early type
galaxies (\cite{faber}). Assuming that the cores are in dynamical equilibrium and supported by random motions, and
that the velocity anisotropy does not vary too much from galaxy to galaxy, and if the core $M/L$ is
a well-behaved function of any two variables $<\mu_b>$, $R_b$, or $\sigma_0$,  then one expects that  the
cores of galaxies follow a law similar to that of the Fundamental Plane (FP).

The main reason to study global  and core properties is that it allows one to probe the mechanism of galaxy formation.
On the other end in the last decade a new ingredient in this picture is represented 
by the discovery of the presence of a massive black bole (BH) in the centers of virtually 
all galaxies (e.g. \cite{FM}, \cite{lauer}).
However, only a small fraction of these BHs exhibit associated nuclear activity 
(non thermal nuclear emission, X-ray and radio emission). The BHs may play an important role in the 
formation and evolution of massive galaxies and are also a key component 
for the development of the nuclear activity. 
Therefore the comparison of the properties of active and inactive galaxies 
through the FP becomes a tool to investigate the interplay between galaxy formation 
and nuclear activity. 

In a previous work, using photometrical and dynamical data for 73 low red-shift (z$<$0.2) 
radio galaxies (RG), we (\cite{bettoni}) were able to compare the FP of RG  with the one 
defined by inactive ellipticals (\cite{jfk96}, JFK96). We showed that 
the same FP holds for both radio and non radio ellipticals with radio galaxies 
occupying the region of the most luminous and large galaxies.

Till very recently few data were available on the optical nuclear properties
of radio galaxies. One of the best optical sets of data is part of the study of
the B2 sample of low luminosity radio galaxies (\cite{fanti}, \cite{Cap00}). For $\sim$60 radio galaxies WFPC2 HST imaging is
 available in the $F555W$ (approximately $V$)  and $F814W$ (approximately $I$) filters. A similar set of data is also available for a sample of powerful 3C radio sources. 
These studies revealed the presence of new and interesting features, some of them
almost exclusively associated to low luminosity FR I radio galaxies.
In particular, the HST observations have shown the presence of dust in a
large fraction of weak (FR I) radio galaxies which takes the form of
extended nuclear disks (\cite{jaffe}, \cite{dekoff}, \cite{dejuan}, \cite{verdoes}, \cite{hans}). 
Such structures have been naturally identified with the reservoir of material which will ultimately
accrete into the central black hole.  The symmetry axis of the nuclear
disk may be a useful indicator of the rotation axis of the central
black hole (see e.g. \cite{capetti1}), although the precise relationship between these two axes remains uncertain.

In this paper we aim to investigate the CFP for a sample of low redshift radio galaxies and to
compare it with that for radio quiet galaxies. The plan of the paper is as follows. In Section 2 we
discuss our observations and data reduction methods. In Section 3 we present the CFP for our sample
of RG. The implications of those observations are then discussed in Section 4. Throughout this paper
we use the Concordance Cosmological Model, with $H_0=70$ kms$^{-1}$Mpc$^{-1}$, and
$\Omega_{\Lambda}=0.70$.


\begin{table*}
\caption{The sample of low redshift B2 radio galaxies with velocity dispersion measurements}             
\label{tab_b2}      
\begin{center}                          
\begin{tabular}{lcccccccc }        
\hline
name & $m_V$ &    z  &    $\sigma_c$ & $\Delta$$\sigma$ &  $<\mu_b>$ &  $\Delta$$<\mu_b>$ & $R_b$  & $\Delta$$r_b$ \\
(1) & (2) & (3) & (4) & (5) & (6) & (7) & (8) & (9) \\
\hline        
0648+27$^+$   &  13.82  &  0.0409 &  137.8  & 38.0 & 14.94 & 0.05 & 0.26 & 0.02  \\                
0755+37    &  13.80  &  0.0413 &  291.0  & 17.3 & 17.05 & 0.15 & 1.09 & 0.09  \\
0908+37    &  15.67  &  0.1040 &  387.1  & 18.0 & 17.36 & 0.10 & 0.50 & 0.05  \\
0915+32B &  15.20  & 0.0620 &  252.1  & 24.3 & 17.00 & 0.05 & 0.52 & 0.03 \\
0924+30    &  13.32  & 0.0266 &  243.7  & 14.5 & 17.33 & 0.15 & 0.96 & 0.14 \\
1003+26    &  15.47  & 0.1165 &  438.4  & 21.6 & 17.94 & 0.15 & 0.41 & 0.05 \\
1113+24    &  15.10  & 0.1021 &  377.8  & 10.2 & 17.44 & 0.05 & 0.47 & 0.03 \\
1204+34    &  15.35  & 0.0788 &  201.0  & 29.6 & 17.47 & 0.25 & 0.46 & 0.09 \\
1322+36B &  12.84  & 0.0175 &  259.1  & 12.1 & 15.45 & 0.05 & 0.62 & 0.05 \\
1339+26B  &  15.10 & 0.0757 &  379.1  & 18.2 & 16.41 & 0.15 & 0.32 & 0.04 \\
1347+28    &  15.66  & 0.0724 &  228.0  & 19.4 & 17.18 & 0.05 & 0.34 & 0.03 \\
1357+28    &  14.81  & 0.0629 &  308.3  & 11.2 & 16.80 & 0.05 & 0.40 & 0.03  \\
1422+26B &  14.32  & 0.0370 &  264.9  & 11.6 & 15.44 & 0.05 & 0.13 & 0.01 \\
1430+25    &  16.65  & 0.0813 &  203.6  & 14.3 & 16.72 & 0.25 & 0.13 & 0.03 \\
1447+27    &  13.88  & 0.0306 &  318.4  & 11.6 & 15.77 & 0.15 & 0.47 & 0.04 \\
1450+28  &  16.31  &  0.1203  & 405.6  & 23.6 & 16.18 & 0.05 & 0.16 & 0.15 \\
1450+28*$^+$  &  16.31   & 0.1265 &  333.3  & 22.1 & 16.56 & 0.05 & 0.20 & 0.01 \\
1502+26   &  15.42   & 0.0540 &  402.1  & 19.9 & 16.60 & 0.15 & 0.66 & 0.07 \\
1512+30   &  15.37   & 0.0931 &  330.2  & 25.6 & 16.55 & 0.05 & 0.34 & 0.03 \\
1521+28   &  15.09   & 0.0825 &  323.0  & 11.8 & 18.02 & 0.15 & 0.85 & 0.06 \\
1527+30   &  15.64   & 0.1143 &  448.0  & 23.0 & 16.76 & 0.10 & 0.34 & 0.03 \\
1553+24   &  14.41   & 0.0426 &  269.9  & 10.9 & 16.41 & 0.05 & 0.45 & 0.03 \\
1557+26   &  14.97   & 0.0442 &  310.7  & 16.1 & 15.45 & 0.05 & 0.21 & 0.01 \\
1613+27   &  14.90   & 0.0647 &  246.6  & 13.5 & 16.74 & 0.15 & 0.30 & 0.03 \\
1658+30   &  14.47   & 0.0351 &  315.8  & 10.2 & 15.02 & 0.15 & 0.19 & 0.02 \\
1726+31   &  17.04   & 0.1670 &  294.8  & 26.8 & 15.96 & 0.15 & 0.22 & 0.03 \\
1827+32   &  15.10   & 0.0659  & 308.5  & 11.4 & 16.24 & 0.05 & 0.21 & 0.01 \\
2236+35   &  12.57   & 0.02759 & 297.8   & 12.5 & 17.09 & 0.05 & 1.28 & 0.11  \\
\hline    
0034+25 &  13.35 &  0.031849   &   295.0   & 10.0 & 16.27 & 0.05 & 0.74 & 0.04 \\
0055+26 &  13.80 &  0.047400   &   231.0   & 13.0 & 16.34 & 0.05 & 0.40 & 0.02 \\
0120+33  & 11.43  &  0.016458   &   315.0   & 10.0 & 15.13 & 0.05 & 0.75 & 0.07 \\
1217+29 &  10.41 &  0.002165   &   238.0   & 10.0 & 13.87 & 0.05 & 0.80 & 0.06 \\
1256+28 &  13.88 &  0.022879   &   203.0   & 10.0 & 16.48 & 0.15 & 0.70 & 0.07 \\
1257+28 &  11.94 &  0.024097   &   278.0   & 10.0 & 17.97 & 0.10 & 2.55 & 0.01 \\
1525+29 &  14.95  &  0.065155   &   250.0   & 30.0 & 16.56 & 0.15 & 0.53 & 0.05 \\
1610+29 &  13.29 &  0.031852   &   331.0   & 26.0 & 16.49 & 0.05 & 0.95 & 0.06 \\
3C88       &  14.24  & 0.030221    &   190.0   & 22.0 & 17.47 & 0.05 & 0.98 & 0.09 \\
3C192     &  15.72  & 0.059709    &   199.0   & 23.0 & 14.34 & 0.15 & 0.09 & 0.01 \\
3C388     &  14.77 & 0.091700    &   365.0   & 23.0 & 18.03 & 0.15 & 0.96 & 0.09 \\
\hline                                   
\end{tabular}
\end{center}
\tiny{Columns: (1) identification of the source, (2) apparent magnitude $m_V$ from \cite{col75} (photographic magnitudes converted to visual magnitudes,
\cite{fan78}, \cite{gon93}, and \cite{gon00}, (3) redshift, 
(4, 5) measured velocity dispersion and error in km/sec, (8 ,9), $<\mu_b>$ and error in mag/$arcsec^2$, (8, 9) break radius $r_b$ and error in arcsecs,  from \cite{hans1};~
 *companion of the radio galaxy, + not used in the CFP see Appendix A1}

\end{table*}
\section{The Sample of radio galaxies}
One of the most complete and well studied samples of nearby radio galaxies, in the northern
hemisphere, is the B2 sample (\cite{fanti}). This sample consists of $\sim$100 early-type galaxies
and was extensively studied at radio wavelengths, especially since the 1980's (see \cite{fanti},
\cite{parma},  \cite{deruiter}, \cite{morg97}). The sample is complete down to 0.25 Jy at 408 MHz
and down to (roughly) $m_v$ = 16.5 and should be essentially unbiased for orientation. The objects
span the radio power range between $10^{23}$ and $10^{26}$ $WHz^{-1}$ at 1.4 GHz with a pronounced peak around $10^{25}$ $WHz^{-1}$.
Therefore they give an excellent representation of the radio source types encountered below and
around the break of the radio luminosity function.

We considered here the sub sample of the B2 sample ( $\sim$ 100 objects) of 57 low redshift (z$<0.20$) radio galaxies, for which photometric and structural data  of the core are available from HST observations in the F814W band (\cite{hans}, \cite{Cap00}). As described in \cite{Cap00} there is no bias in the selection of the 57 objects observed with HST (they were chosen randomly as far as their radio and optical properties are concerned).  In comparing  various parameters of this observed sub-sample with those 
of the sources that were not observed by HST they found that no significant differences emerged. The HST observations are complete at the level of $\sim$57\% and therefore constitute an unbiased sub sample.

\section{Summary of  previous results on the core properties of radio galaxies}

In \cite{hans} the dust properties of the radio cores were analysed. About half of the sources have
significant amounts of dust in the nuclear region, mostly in the form of disks or lanes, and if
radio jets are present they tend to be perpendicular to the dusty disk or lane, at least in the low
power sources ($P < 10^{24}$ WHz$^{-1}$ at 1.4 GHz). There is also a (broad) correlation between
total dust mass of the disks and the radio power.

Based on high resolution HST images it was found that elliptical galaxies come in two flavours, one with "core" radial brightness profiles and one with power law profiles (see e.g. \cite{faber}). This distinction is seen also for other properties: rotation, isophote shape and presence of X-ray emission (see for a recent discussion \cite{kormendy09}).

This dichotomy extends also to radio properties. Radial brightness profiles of sources with moderate content of nuclear dust (\cite{hans}) are of core type. This as has been confirmed by \cite{bc06} and \cite{cb06}. The observations suggest that there is a class of early-type galaxies that will never harbour a classical radio source of the Fanaroff-Riley type I or II.
These galaxies have steep power law inner profiles, with a slope (inner Nuker law) $\gamma > $0.3. On the other hand galaxies with a core profile ($\gamma < $0.3) may or may not possess a nuclear radio source. This difference seems genuine and not induced by selection effects. In fact the absolute magnitudes of core and power law galaxies overlap and the power law galaxies are never associated with classical radio sources.
This result is consistent with the dichotomy described above and suggest that AGN activity 
in core galaxies can be very pronounced and that the dichotomy is the result of strong or weak AGN feedback respectively (\cite{kormendy09}).

\section{Stellar velocity dispersion}

To derive the CFP we need, together with the photometric data, the central velocity dispersion of
the galaxy. For 11 galaxies in the sample we found measurements of the stellar velocity dispersion from the
literature, using the Lyon-Meudon Extragalactic Database (LEDA \cite{paturel}). These measurements
of $\sigma$ were corrected to a circular aperture with a metric diameter of 1.19 $h^{-1}$kpc ,
equivalent to 3.4$^{\prime\prime}$ at the distance of the Coma cluster. To derive  $\sigma_0$ for
the remaining of the sources we carried out spectroscopic observations. We were able to observe 27
out of the remaining 46 galaxies. One object (0648+27) was not included in the final sample to derive the CFP, as explained in Appendix A1. For this reason the final sample considered for this study is therefore composed of 37 objects.

\subsection{Spectroscopic observations}

Optical spectroscopy in the range $\lambda\lambda$=3700-4700 \AA ~and 4600-6500 \AA
was obtained in service mode in two observing runs in 2005 and 2006 with the
Telescopio Nazionale Galileo (TNG). We used the spectrograph DOLORES 
equipped with a Loral CCD with $2048 \times 2048$ pixels of 15~$\mu$. 
In 2005 we used the HRV Grism \#6 with a slit of 1.0 arcsec, this yields a velocity dispersion
resolution $\Delta \sigma$=75 km$s^{-1}$. In 2006 we used the grism VHR-V with a slit of 1.0 arcsec,
this yields a $\Delta \sigma$=85 km$s^{-1}$. The plate scale across the dispersion is 0.275 arcsec/
pix. In addition to the radio galaxy spectra, we secured spectra of bright stars of spectral type
from G8III to K1III with low rotational velocity, (V$\times$$seni<$17 km$s^{-1}$). These spectra
were used as templates of zero velocity dispersion. During the second observing run also an early-
type galaxy (NGC 3377), as standard for the measure of $\sigma$, was observed. The slit was oriented
along the apparent major axis for all the galaxies in our sample except in the cases of galaxies
with multiple nuclei, where the slit was aligned along the two nuclei. All the spectra were bias and
flat-field corrected, trimmed and wavelength calibrated using standard procedures available in the
IRAF package. The accuracy of the latter procedure was checked with measurements of the night sky
$\lambda_0$ = 5577.32 \AA~ emission line. The systemic velocity, corrected to the Sun, and the
velocity dispersion $\sigma$ were determined using the Fourier Quotient method (\cite{sargent};
\cite{bertola}).   The Fourier Quotient method was applied, using all the template stars, to all
spectra to obtain the radial velocity and the velocity dispersion. The rms of the determinations
obtained with different template stars turned out to be less than 20 $km s^{-1}$ for $\sigma$ and
$\sim10 ~km s^{-1}$ for the systemic radial velocity $V_r$ both for the sample of radio galaxies
than for the template galaxy. For NGC 3377 we found $\sigma$=160$\pm25$ km$s^{-1}$ in agreement with
the value of 139  km$s^{-1}$ reported in LEDA. The average values of individual determinations were
adopted as final values of $\sigma$ and $V_r$. In order to improve the S/N ratio of the spectra we
co-added the central spectra within an aperture of 6$^{\prime\prime}$. Since early-type galaxies
exhibit some gradients in the radial velocity and velocity dispersion, the derived central parameter $\sigma$ depends on the distance of the galaxies and on the size of the aperture used for the observation. In order to compare our velocity dispersions with the data available in the literature we applied aperture corrections according to the procedure given by \cite{jfk96}. The individual measurements of $\sigma$ are corrected to a circular aperture with a metric diameter of 1.19 $h^{-1}$kpc , equivalent to 3.4$^{\prime\prime}$ 
at the distance of the Coma cluster of galaxies.
\begin{figure}
   \centering
   \includegraphics[height=9.5cm]{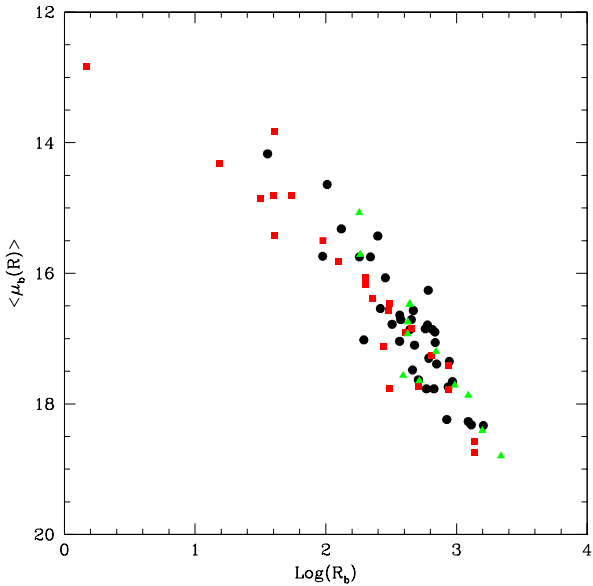}
      \caption{The $<\mu_b>$--$R_b$ relation of B2 radio galaxies  ([black] circles) and 
      nearby (non radio) early type galaxies ([red] squares
\cite{faber}, [green] triangles \cite{bernardi}).}
         \label{mubrb}
   \end{figure}  
\begin{figure}
  \centering
     \includegraphics[height=9.5cm]{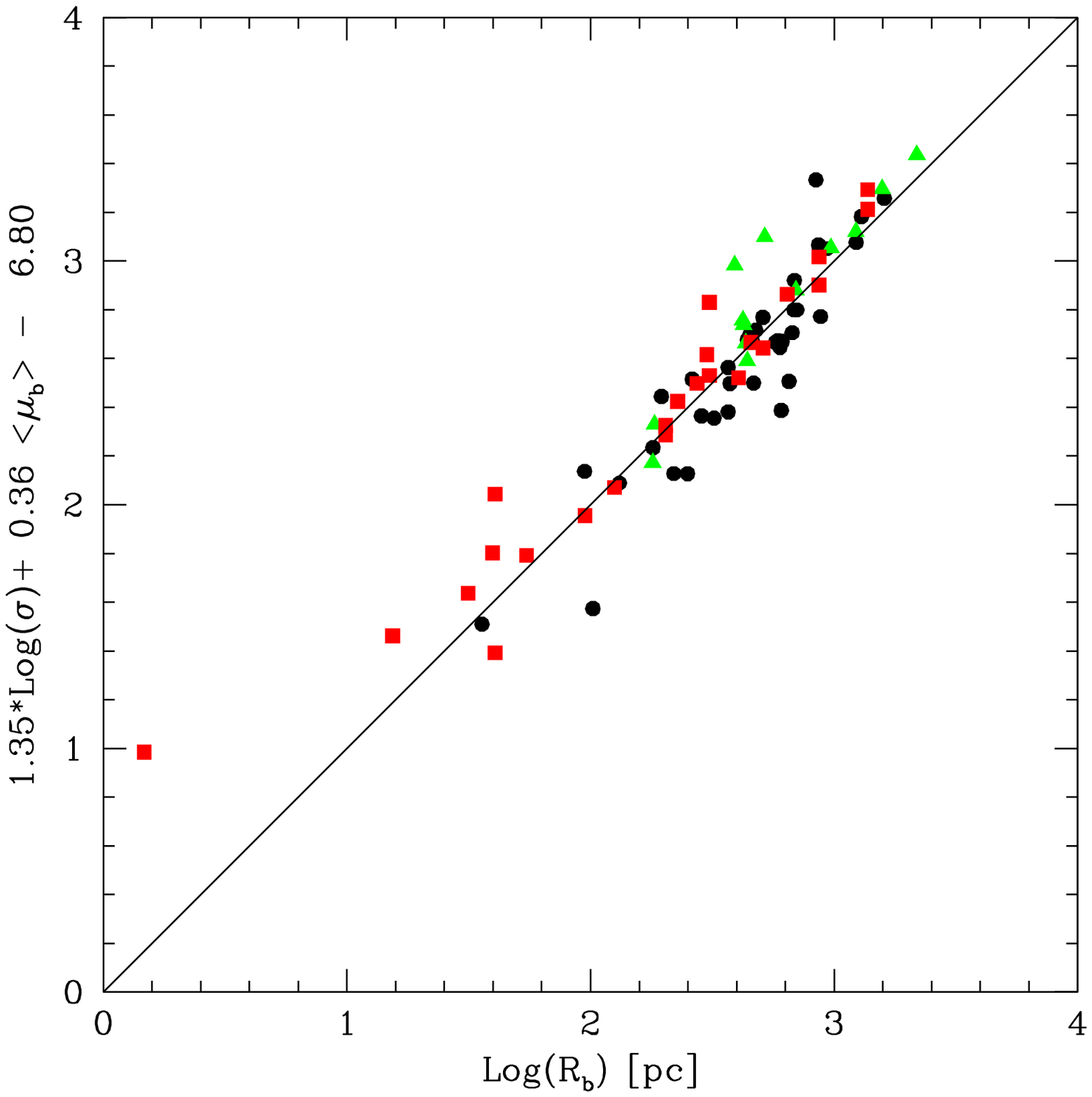}
     \includegraphics[height=9.5cm]{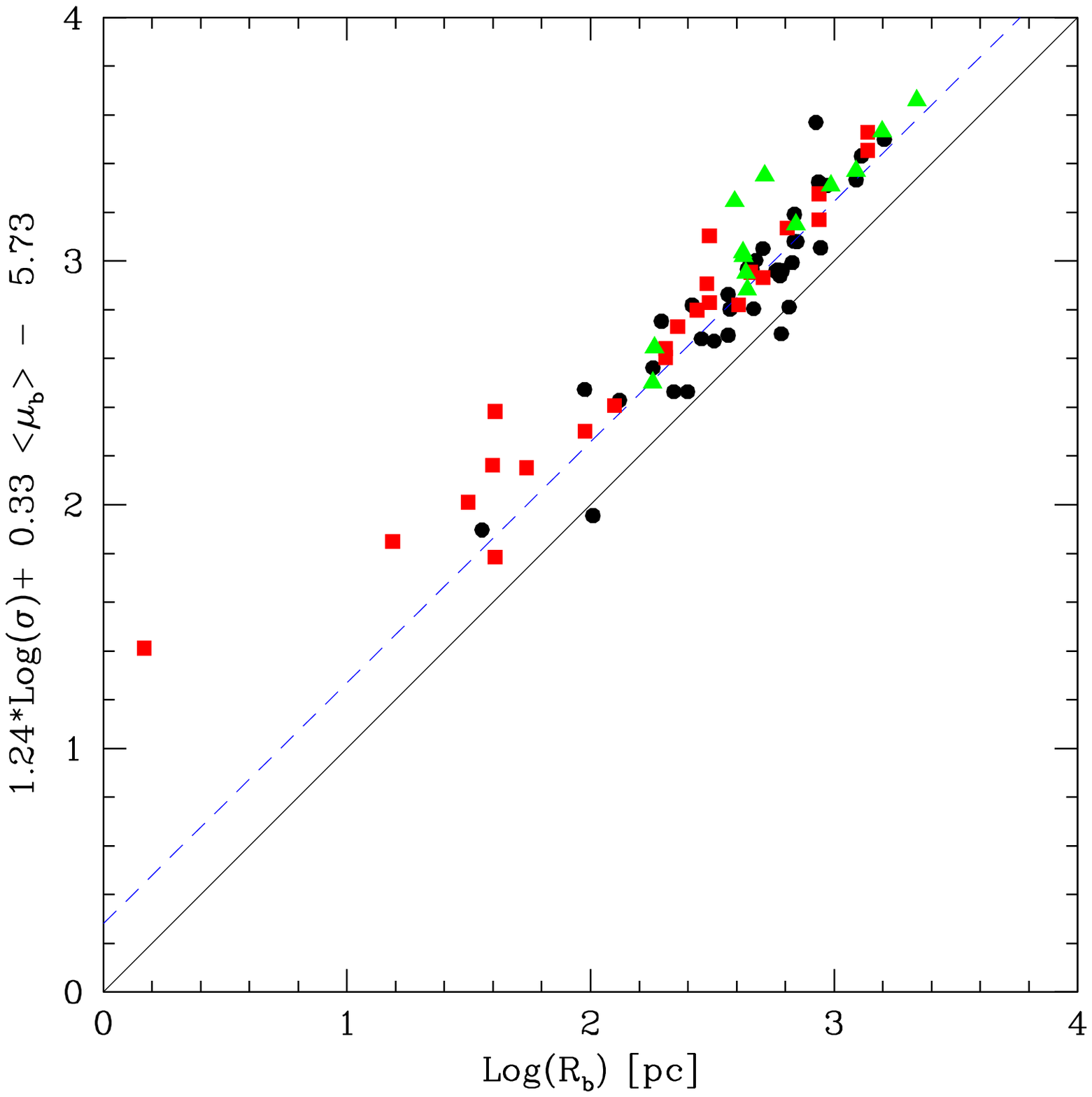}
          \caption{{\it Upper panel:} The best fit of the CFP for the B2 radio galaxies 
          (black circles) 
compared with a sample of nearby (non radio) early type galaxies ([red] squares
\cite{faber}, [green] triangles \cite{bernardi}), using the CFP fit in the y-axis. 
The solid line represents CFP relation.
Lower panel: The same data as in the upper panel, but using the coefficient of the global 
FP fit of (\cite{bettoni}) with a shift in the y-axis to take into account the different units in the x-axis
(from kpc to pc) ; the solid line is the corresponding FP relation. The blue dashed line report the fit to
the core data with a shift of 0.23dex}
         \label{CFP}
   \end{figure}  


\begin{figure}
   \centering
   \includegraphics[height=9.5cm]{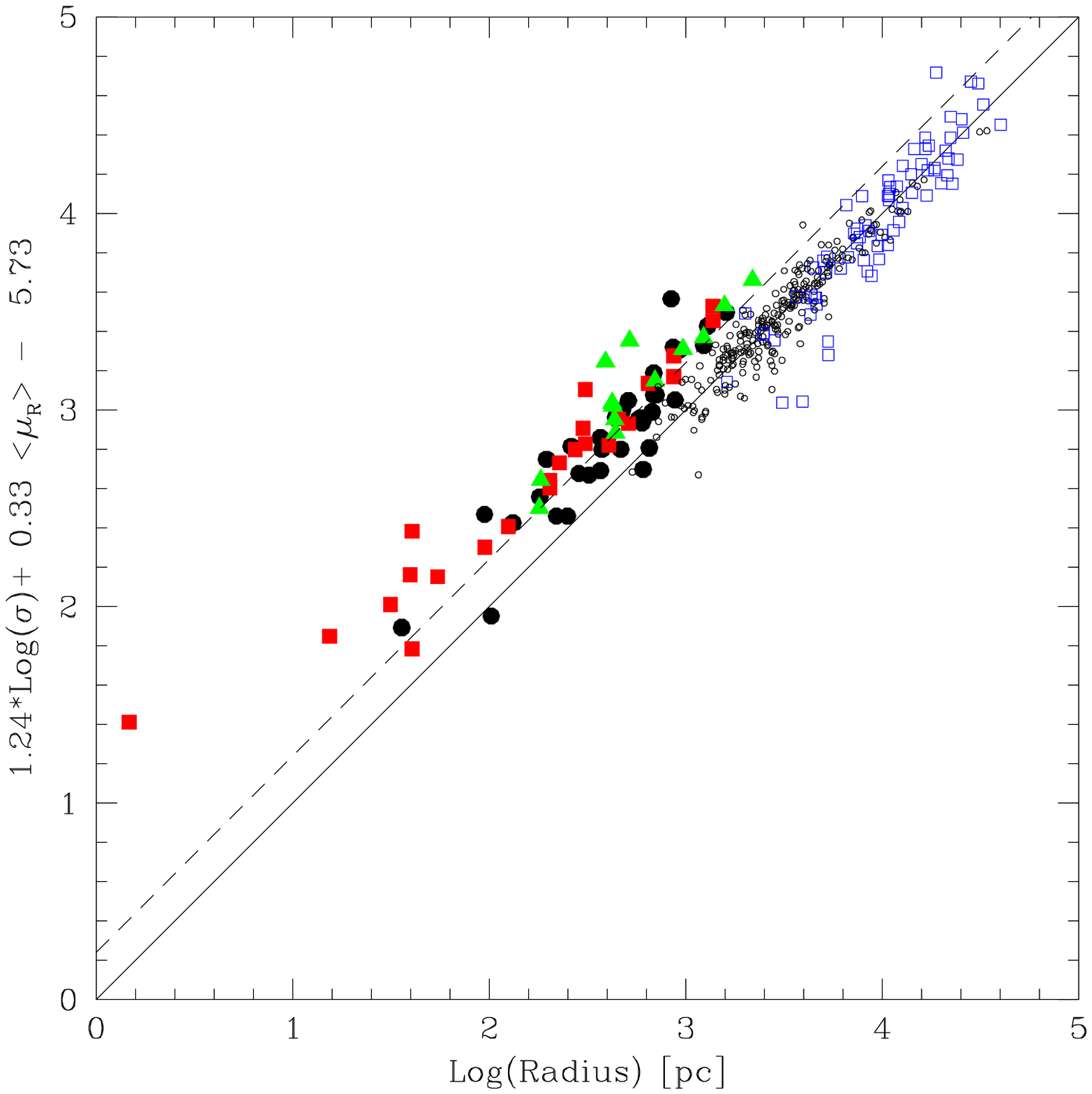}
      \caption{The CFP of B2 radio galaxies  (black circles) and 
      nearby (non radio) early type galaxies ([red] squares
\cite{faber}, [green] triangles \cite{bernardi}) 
      compared to the Fundamental Plane of normal (small black dots)\cite{jfk96} and
radio galaxies light-blue squares (\cite{bettoni}). The solid line represents the FP for
a sample of RG (\cite{bettoni}) while the dashed 
line yields the fit to the core data.  Note that we plot $R_b$ and$<\mu_b>$ for the CFP data (black circles, red squares and green triangles) and $R_e$ and $<\mu_e>$ for the  FP data (small black open circles and small blue open squares).}
         \label{CFP1}
   \end{figure}  

  All these data are listed in Table \ref{tab_b2}, in which we also give
photometric data derived from HST images in the F814W filter (all taken with exposure time of 300 seconds): as described in (\cite{hans}) the one-dimensional brightness profiles determined with IRAF task ELLIPSE were fitted with a Nuker-law, thus obtaining
for each galaxy, in the F814W (I) filter, the core radius $R_b$ in arcsecs and the average surface brightness $<\mu_b>$ (in mag/arcsecs$^2$) within the core radius.  

For some galaxies the S/N is enough to study the inner part of the rotation curve (see Fig \ref{rot}). In Appendix A comments on some individual objects are reported.

\section{The CFP of radio and normal ellipticals}
To complete our data-set we decided to collect all the available data in literature for the core properties of E galaxies. We added to the RG data the core photometric data and the velocity dispersion for two sets of data of non radio galaxies. 
The first set is from the classical work of \cite{faber} from which we have data for 56 normal 
galaxies. To compare these measurements with our data for RG (derived in the R band) we applied a 
mean color correction $V-R=0.60$. The second set is taken from the  
recently published HST photometric data (\cite{hyde}) and  the corresponding central 
velocity dispersion $\sigma$ (\cite{bernardi}) for the core of 13 nearby non radio massive galaxies.

The data of \cite{hyde} are obtained using the High Resolution Channel (HRC) of the ACS in the i
band, thus we transformed their CTPS to mag/sq/arcs using the zero point for the i filter $ZP$ =
25.654 (AB mag). Then we applied a correction $\Delta$m= -0.53 to transform the data to Vega
magnitude system. In figure \ref{mubrb} we plot the $<\mu_b>$--$R_b$ projection of the CFP for our sample of radio galaxies and for a combined sample of non radio ellipticals (\cite{faber}, \cite{bernardi}), we confirm the same correlation already found by \cite{faber}. 

In order to derive the parameters ($\alpha$, $\beta$ and $\gamma$) describing the CFP in the following relation: 

\begin{equation} 
\log R_b = \alpha\log \sigma + \beta<\mu_b> + \gamma , 
\end{equation} 

\noindent 
we minimized the root square of the residual perpendicular to the plane.
In Table~\ref{FIT} we report the coefficients obtained using this fitting procedure for the CFP of both radio galaxies and normal ones,  for comparison we report also the coefficients for the FP for radio and normal galaxies (\cite{bettoni}).

\begin{table}
\centering
\caption{Coefficients of the best fit for the Core Fundamental Plane and 
the global Fundamental Plane of low redshift radio galaxies (see eq. 1 ) and of normal galaxies.}             
\label{FIT}                              
\begin{tabular}{c c c c | c c c c }        
\multicolumn{1}{l}{}& \multicolumn{3}{c}{Radio Galaxies} & \multicolumn{3}{c}{Normal Galaxies} \\
\hline\hline                
 & $\alpha$ & $\beta$ & $\gamma$ & $\alpha$ & $\beta$ & $\gamma$\\    
\hline                        
   CFP & 1.35 & 0.36 & -6.80 & 1.32 & 0.36 & -6.75 \\      
   FP  & 1.24 & 0.33 & -5.96 & 1.27 & 0.32 & -5.58 \\
\hline                                   
\end{tabular}
\end{table}		
		
In Figure \ref{CFP} we show the comparison of the CFP plotted with the coefficients of our best fit and
with the coefficients of the best fit of FP (\cite{bettoni}). The two planes are very similar and almost
parallel (see also Tab. 2). We note a slight tendency of CFP to show a curvature at low luminosity; this is similar to what pointed out by \cite{desro07} for the FP.

In Figure \ref{CFP1}  we compare the whole CFP with the FP derived from the global properties of a
different sample of radio and non radio ellipticals (see \cite {bettoni} for details). Again the CFP and FP are
closely parallel as suggested by \cite{faber}: whatever differences exist among core
galaxies these are not so large as to erase the appearance of a two-parameter family
of self-gravitating cores that is fundamentally not too dissimilar from the
2-dimensional family of isothermal spheres.

\section{Conclusions}

We have presented the photometric, structural and kinematic properties  of the centers of a sample
of 38 low redshift radio galaxies galaxies and have shown that the conventional parameters
characterizing the centers (the break radius, its surface brightness and the central velocity
dispersion) are well represented in a plane (the Core Fundamental Plane) which is indistinguishable
from that of elliptical galaxies that do not exhibit radio emission.

A similar result was found from the comparison of the  global properties of a sample of 72  radio
galaxies and local Ellipticals using the standard Fundamental Plane description (\cite{bettoni}).
The comparison of the properties of the FP, that refer to the whole galaxy, with those concering the
centers (the CFP) shows that the slopes of the two  planes are very similar and suggest taht the same
mechanism is responsible for the link among the involved quantities.

The remarkable similarity of the properties of radio and non radio elliptical galaxies in the 
description of both the Fundamental Plane and the Core Fundamental Plane indicates that the active phase 
of the galaxy connected with the strong emission at radio frequencies has likely an inconsequential
effect for the structure of the whole galaxy.  Moreover these results suggest that the two type of galaxies 
(radio and non radio) have had a similar history of formation and evolution.

 \appendix\section{Notes to individual galaxies}

\begin{figure*}
   \centering
   \includegraphics[angle=-90,width=7.5cm]{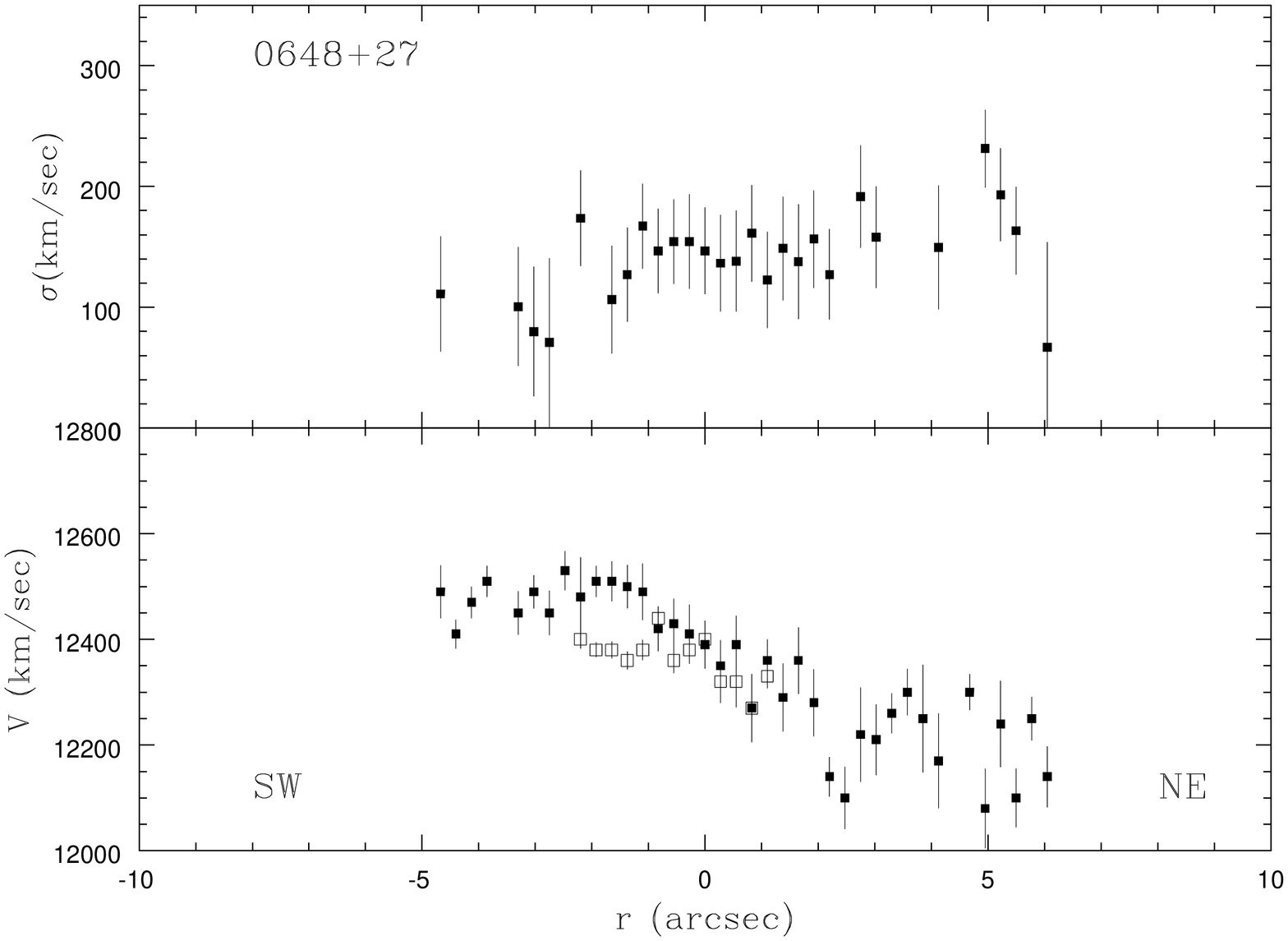}
      \includegraphics[angle=-90,width=7.5cm]{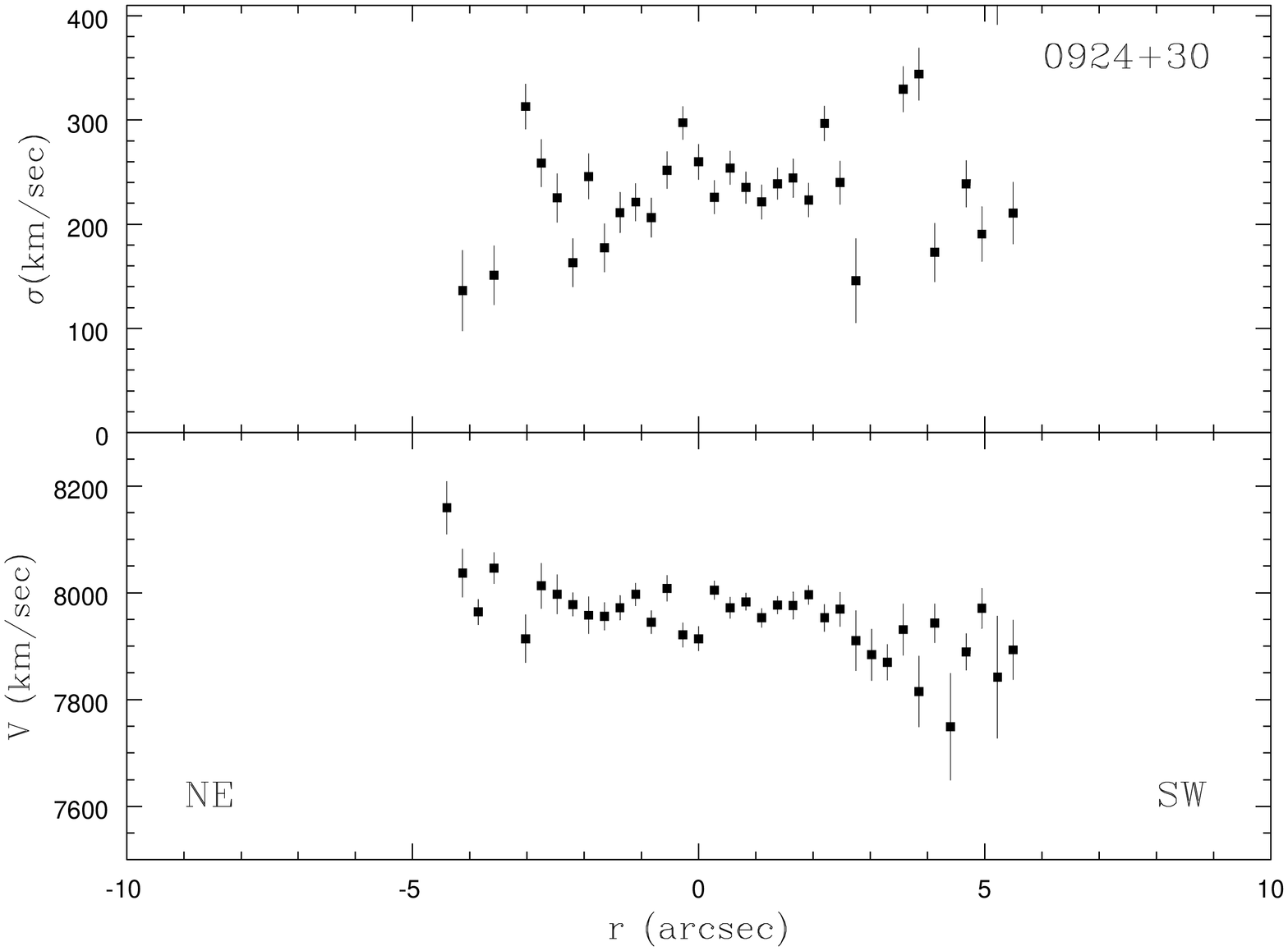}
   \includegraphics[angle=-90,width=7.5cm]{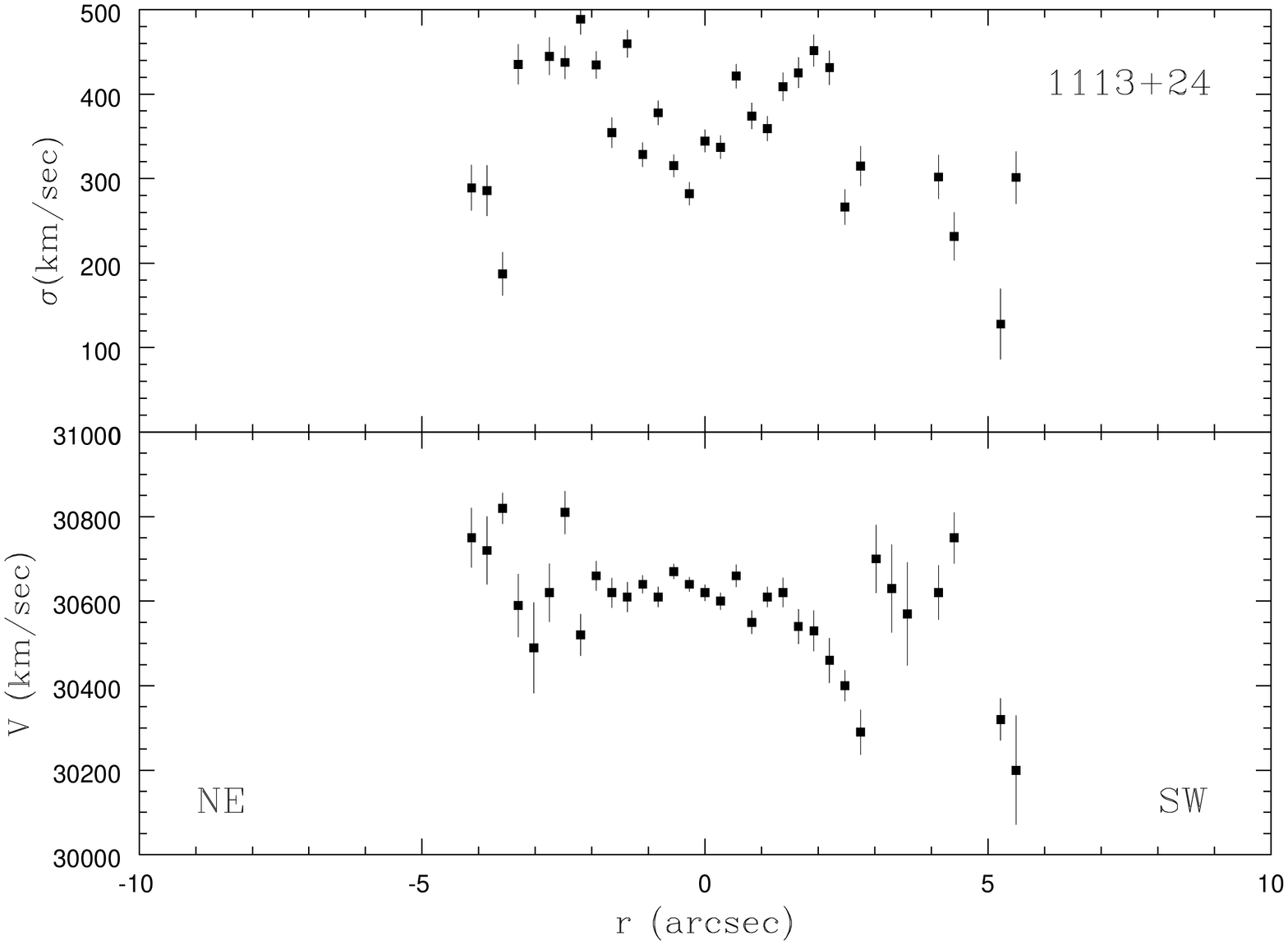}
         \includegraphics[angle=-90,width=7.5cm]{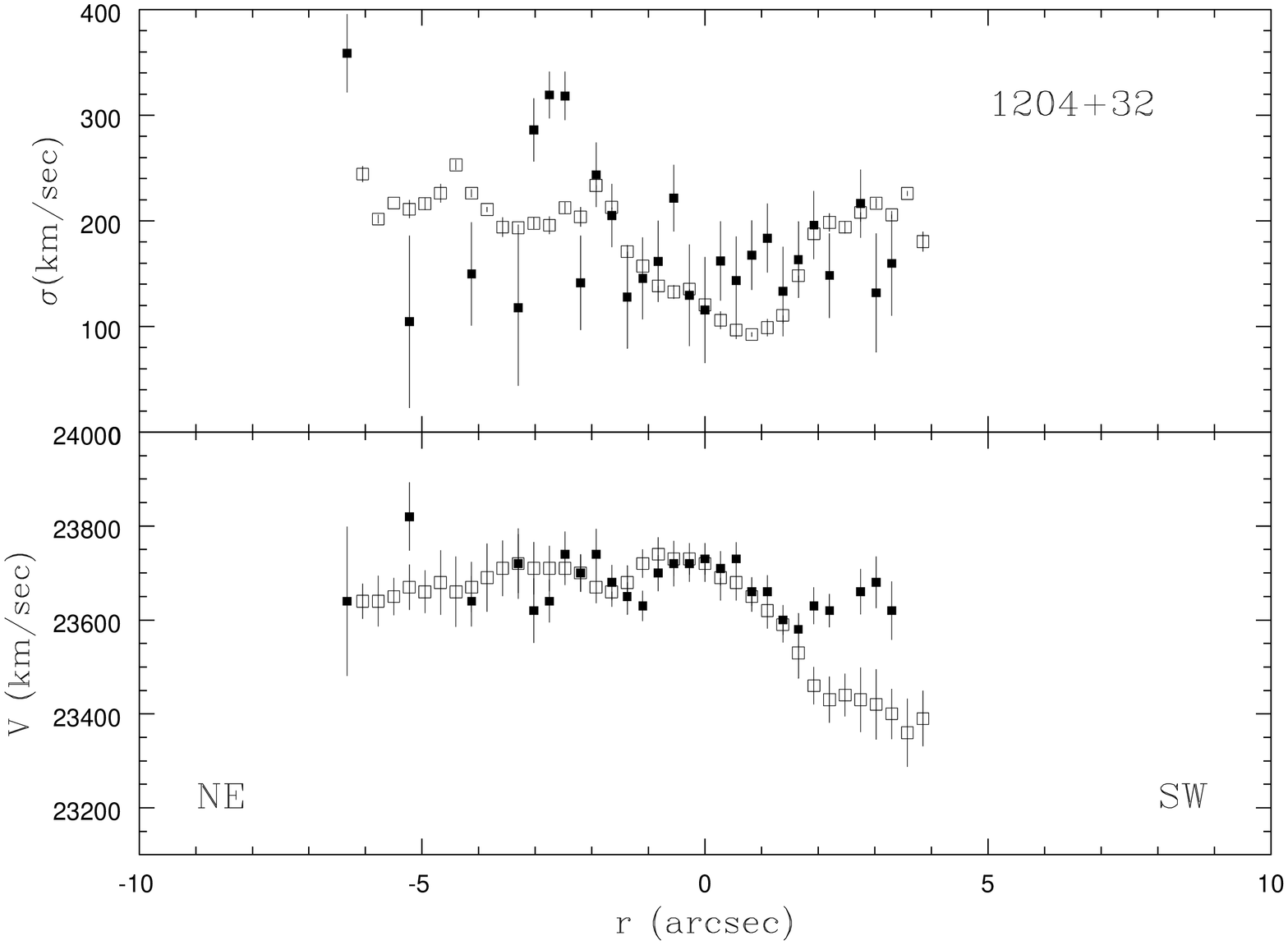}  
   \includegraphics[angle=-90,width=7.5cm]{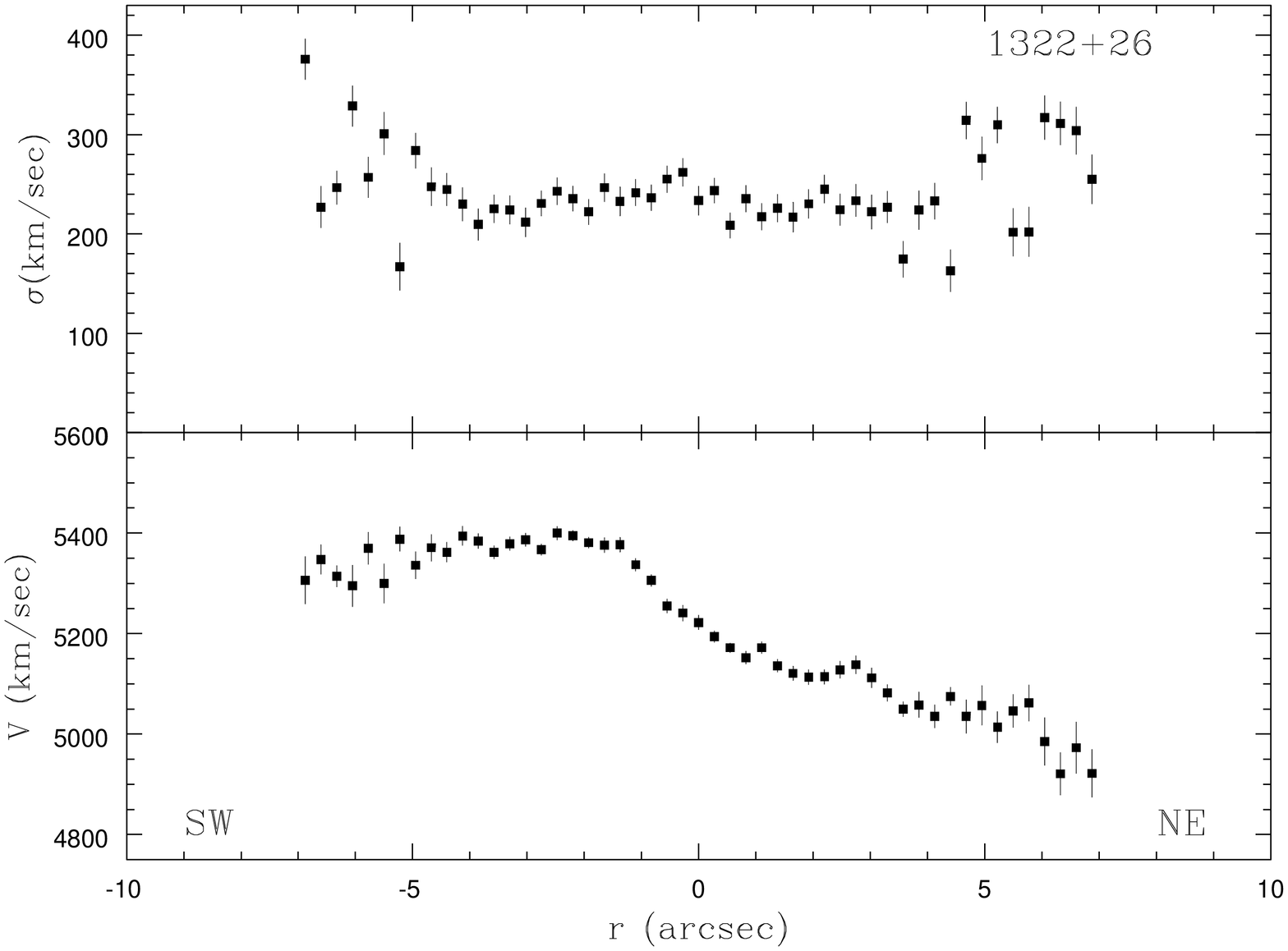}
       \includegraphics[angle=-90,width=7.5cm]{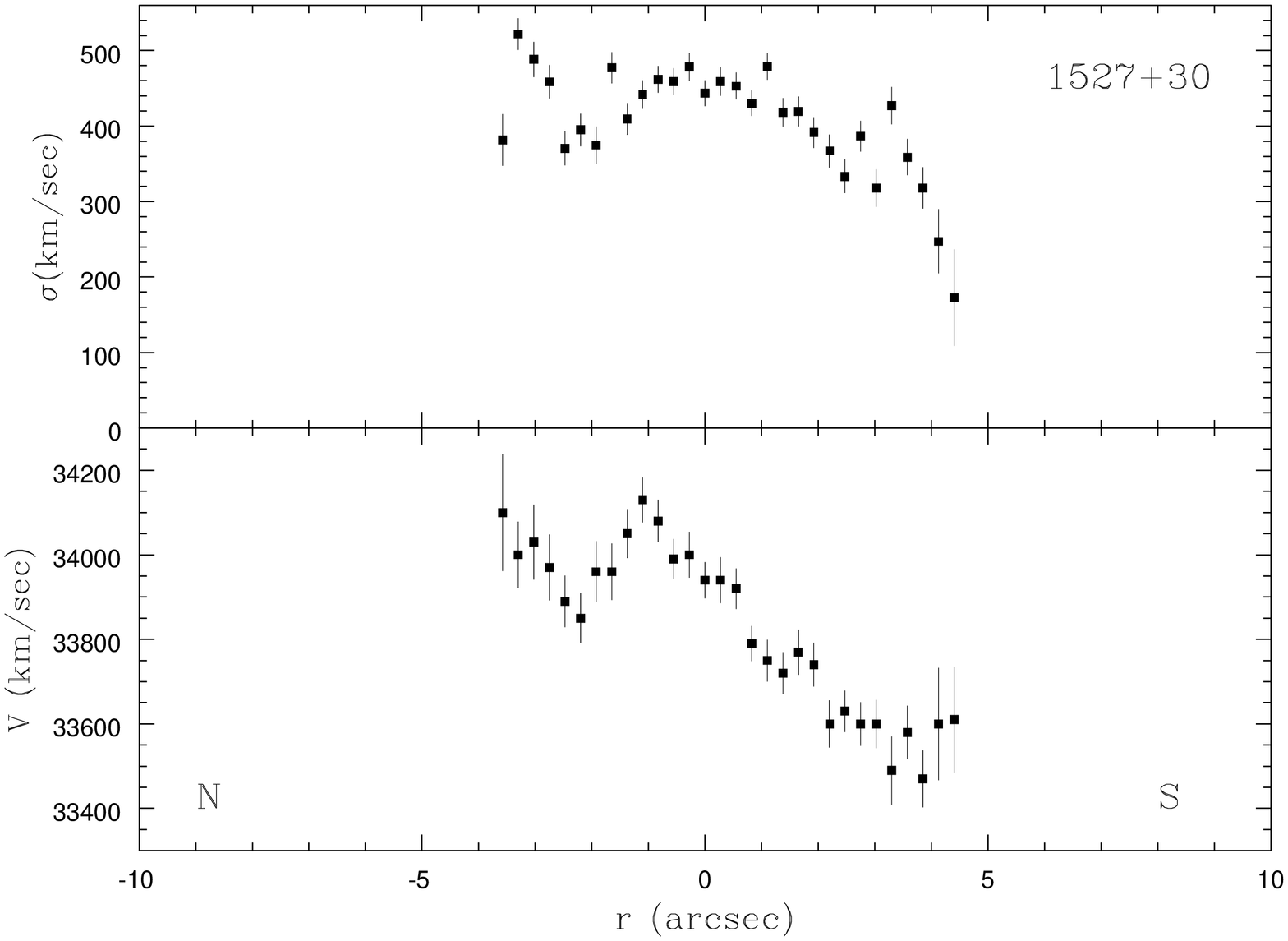}
      \caption{Rotation curves and velocity dispersion profiles for 
      selectd low redshift radio galaxies (full squares stars, open squares gas). }
         \label{rot}
   \end{figure*}

B2 0648+27 - This galaxy show a typical E+A spectrum indicating the presence of a young stellar population (\cite{emonts1}, \cite{emonts2}). 
In this last case the measured velocity dispersion is only indicative being measured using a not perfect stellar template. 
This galaxy was not used to construct the CFP of RG. The spectrum show the [OII] $\lambda$3727\AA~ emission line. 
For this object we were able to measure the rotation curve (both for gas and stars) and the velocity dispersion profile  in the inner  $\sim10^{\prime\prime}$ 
these are shown in figure \ref{rot}. We found a $V_{max}$=200 km/sec for both components. 

B2 0908+37- As noted by \cite{Cap00} a fainter companion galaxy is located $\sim3.7^{\prime\prime}$ to the SW. 
Our spectrum cross also the faint companion for which we were able to measure the redshift $z_{comp}$=0.1027 very close to the redshift of the radio galaxy.

B2 0924+30 - For this galaxy we measure a rotation $V_{max}$=100 km/sec in the inner $\sim5^{\prime\prime}$ (see figure \ref{rot}).

B2 1113+24 - For this galaxy we measure a rotation $V_{max}$=180 km/sec in the inner $\sim5^{\prime\prime}$ (see figure \ref{rot}). 

B2 1204+34 - This galaxy show strong emission lines in the spectrum i.e [O III] $\lambda$ 5007~ \AA~ and $H_{\beta}$. 
We measured the gas rotation curve, and after subtraction of the emission component also the stellar rotation curve (see figure \ref{rot}). 
The gas is more extended in the NE direction and show a $V_{max}\sim$200 km/sec.

B2 1322+36 - NGC 5141 - This S0 galaxy has a nuclear dust lane and the gas exhibits a regular rotation profile as measured by  \cite{nsj}. 
Here we measure, for the first time also a regular rotation curve for the stellar component with a $V_{max}$=200 km/sec at r=$\sim10^{\prime\prime}$ (see figure \ref{rot}). 
Our spectra are in the region $\lambda$ 3800-4800 \AA~ were only a very faint $\lambda$3727 [OII] line is visible, the gas and stellar component are co-rotating.

B2 1422+26B - This galaxy show the presence in the spectrum of the [O II] $\lambda\lambda$ 3727 \AA~ line

B2 1450+28 - This galaxy is a well known dumbbell system in the center of the cluster Abell 1984. The radio
galaxy is associated with the southern galaxy of the pair. To obtain our spectrum, the slit was oriented at PA= $0^{\circ}$ and cross both galaxies. 
In Table \ref{tab_b2} we report the velocity dispersion for both objects. In table \ref{tab_b2} the non radio galaxy is labelled with an asterisk and was excluded from the fit. 

B2 1502+26  - 3C310 This galaxy show a boxy elliptical galaxy approximately $15^{\prime\prime}$ SW, the slit for our spectrum was oriented at 
PA=$90^{\circ}$ and cross both galaxies, for the fainter companion we measure the same redshift z=0.0540 and a velocity dispersion $\sigma$= 390.0$\pm$ 22.7 km/sec.


B2 1527+30 - in Abell 2083; this galaxy show a strong velocity gradient $V_{max}$=180 km/sec in the inner $10^{\prime\prime}$ (see figure \ref{rot}). 

\begin{acknowledgements}
We thanks the anonymous referee whose comments helped us to improve our manuscript. This research made use of Vizier service (\cite{Ochsenbein}) and
of the NASA/IPAC Extragalactic Database (NED) which is operated by the
Jet Propulsion Laboratory, California Institute of Technology, under
contract with the National Aeronautics and Space Administration. We
have made use of the LEDA (http://leda.univ-lyon1.fr) and Hypercat
database.      
\end{acknowledgements}

\end{document}